# DNS Study for the origin of the flow Randomization in Late Boundary Layer Transition

Manoj Thapa[1], Ping Lu[2] and Chaoqun Liu[3]

University of Texas at Arlington, Arlington, Texas 76019, USA

cliu@uta.edu

## Abstract

This paper is devoted to the investigation of the origin and mechanism of randomization in late boundary layer transition over a flat plate without pressure gradient. The flow randomization is a crucial phase before flow transition to the turbulent state. According to existing literatures, the randomization was caused by the big background noises and non-periodic spanwise boundary conditions. It was assumed that the large ring structure is affected by background noises first, and then the change of large ring structure affects the small length scales quickly, which directly leads to randomization and formation of turbulence. However, by careful analysis of our high order DNS results, we believe that the internal instability of multiple ring cycles structure is the main reason. What we observed is that randomization begins when the third cycle overlaps the first and second cycles. A significant asymmetric phenomenon is originated from the second cycle in the middle of both streamwise and spanwise directions. More technically, a visible asymmetric phenomenon in the middle vortex ring cycle starts at time step t=16.25T and x=838.9δin where the top and bottom level rings are still completely symmetric. The non-symmetric structure of middle level ring affects the small length scale in boundary layer bottom quickly. The randomization phenomenon spreads to top level through ejections. Finally, the whole flow domain becomes randomized. A hypothesis of C- and K-types shift is given as a possible mechanism of flow randomization.

## Nomenclature

$M_\infty$ = Mach number  
$Re$ = Reynolds number  
$\delta_{in}$ = inflow displacement thickness  
$T_w$ = wall temperature  
$T_\infty$ = free stream temperature  
$Lz_{in}$ = height at inflow boundary  
$Lz_{out}$ = height at outflow boundary  
$Lx$ = length of computational domain along x direction  
$Ly$ = length of computational domain along y direction  
$x_{in}$ = distance between leading edge of flat plate and upstream boundary of computational domain  
$A_{2d}$ = amplitude of 2D inlet disturbance  
$A_{3d}$ = amplitude of 3D inlet disturbance  
$\omega$ = frequency of inlet disturbance  
$\alpha_{2d}, \alpha_{3d}$ = two and three dimensional streamwise wave number of inlet disturbance  
$\beta$ = spanwise wave number of inlet disturbance  
$R$ = ideal gas constant

---

[1] PhD Student, AIAA Student Member, University of Texas at Arlington, USA
[2] PhD Student, AIAA Student Member, University of Texas at Arlington, USA
[3] Professor, AIAA Associate Fellow, University of Texas at Arlington, USA, AIAA Associate Fellow

$= $ ratio of specific heats $\qquad\qquad \mu_\infty =$ viscosity

## I.   Introduction

Turbulence, one of the top secrets of the nature, in general, is composed of two parts:  small length scale generation and flow randomization . Although the turbulence formation in late boundary layer transition has become subject of intense study for over a century, there are only few research papers about the mechanism of randomization in late boundary layer transition. One main reason may be that fluid dynamic community always relied on the classical theory at least for turbulence generation and sustenance. According to the classical theory, turbulence is generated when the large vortices in late stage of boundary layer transition break down into small length scales [19]. Unfortunately, breakdown and reconnection process couldn't be observed by our DNS. Moreover, this phenomenon is theoretically impossible and could never happen in practice. For detail [20]

While taking into account earlier research work about flow randomization, one of  the milestone work was done by Daniel Meyer and his colleagues. They believed that "the inclined high-shear layer between the legs of the vortex exhibits increasing phase jitter (i.e. randomization) starting from its tip towards the wall region".[2,3]. However from our numerical simulation, we observed a phenomenon which is different from the theory given by Meyer and his co-workers. We use periodic boundary condition in spanwise direction and disturbances are present only at inflow, outflow and far field. Still, we observe randomized flow. So, it is unlikely to happen randomization due to back ground noise and use of non periodic condition in spanwise direction.

In general, turbulence flow is characterized by chaotic and stochastic property changes in time and space. Then obviously some of important questions may come to our mind such as from where random fluctuation starts and by which mechanism flow becomes randomized? We thoroughly investigate our DNS results and try to trace as earliest as possible location and time step. We also try to reveal new mechanism. Since, skin friction in turbulence boundary layer region is always higher than in laminar region so the people who are dealing with design of aircraft always willing to delay transition process in boundary layer (reducing skin friction in boundary layer of aircraft reduces consumption of fuel). In this context, our present paper may obtain great attention.

A $\lambda_2$-eigenvalue technology developed by Jeong and Hussain (1995) is used for visualization of vortex structures formed by interaction of non-linear evolution of T-S wave in transition process.

## II. Case Setup and Code Validation

### 2.1 Case setup

The computational domain is displayed in Figure 1. The grid level is 1920×128×241, representing the number of grids in streamwise ($x$), spanwise ($y$), and wall normal ($z$) directions.  The grid is stretched in the normal direction and uniform in the streamwise and spanwise directions. The length of the first grid interval in the normal direction at the entrance is found to be 0.43 in wall units (Y$^+$=0.43). The parallel computation is accomplished through the Message Passing Interface (MPI) together with domain decomposition in the streamwise direction (Figure 2). The flow parameters, including Mach number, Reynolds number, etc are listed in Table 1. Here, $x_{in}$ represents the distance between leading edge and inlet, $Lx$ , $Ly$ , $Lz_{in}$ are the lengths of the computational domain in x-, y-, and z-directions, respectively, and $T_w$ is the wall temperature.

Table 1: Flow parameters

| $M_\infty$ | $Re$ | $x_{in}$ | $Lx$ | $Ly$ | $Lz_{in}$ | $T_w$ | $T_\infty$ |
|---|---|---|---|---|---|---|---|

| 0.5 | 1000 | 300.79 $_{in}$ | 798.03 $_{in}$ | 22 $_{in}$ | 40 $_{in}$ | 273.15K | 273.15K |
|---|---|---|---|---|---|---|---|

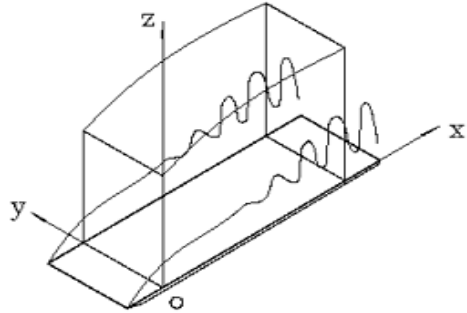

Figure 1: Computation domain

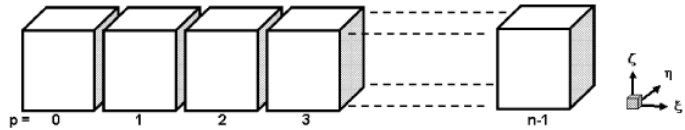

Figure 2: Domain decomposition along the streamwise direction in the computational space

## 2.2 Code Validation

The DNS code – "DNSUTA" has been validated by NASA Langley and UTA researchers (Jiang et al, 2003; Liu et al, 2010a; Lu et al 2011b) carefully to make sure the DNS results are correct.

### 2.2.1 Comparison with Linear Theory

Figure 3 compares the velocity profile of the T-S wave given by our DNS results to linear theory. Figure 4 is a comparison of the perturbation amplification rate between DNS and LST. The agreement between linear theory and our numerical results is quite good.

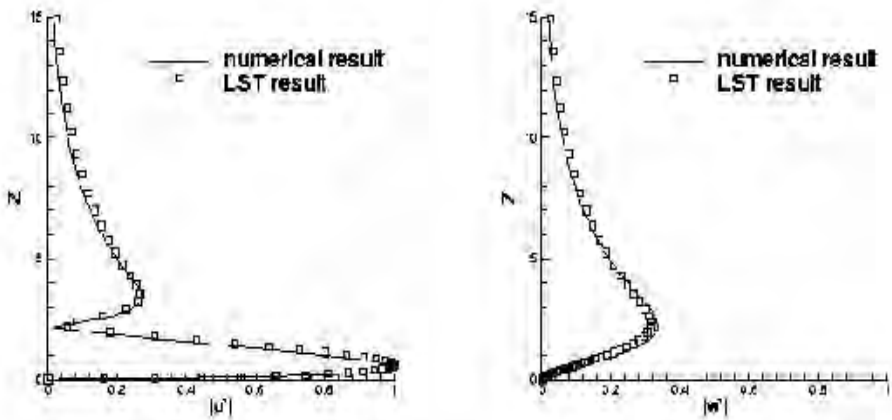

Figure 3: Comparison of the numerical and LST velocity profiles at Rex=394300

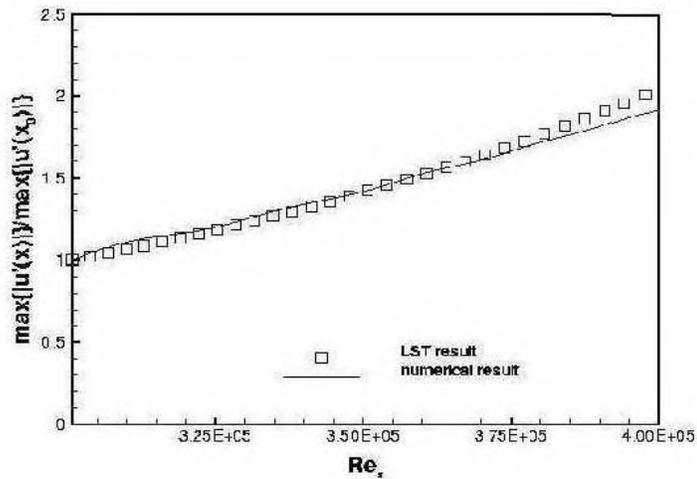

Figure 4: Comparison of the perturbation amplification rate between DNS and LST

### 2.2.2 Grid Convergence

The skin friction coefficient calculated from the time-averaged and spanwise-averaged profile on a coarse and fine grid is displayed in Figure 5. The spatial evolution of skin friction coefficients of laminar flow is also plotted out for comparison. It is observed from these figures that the sharp growth of the skin-friction coefficient occurs after x≈450$\delta_{in}$,which is defined as the 'onset point'. The skin friction coefficient after transition is in good agreement with the flat-plate theory of turbulent boundary layer by Cousteix in 1989 (Ducros, 1996). Figures 5(a) and 5(b) also show that we get grid convergence in skin friction coefficients.

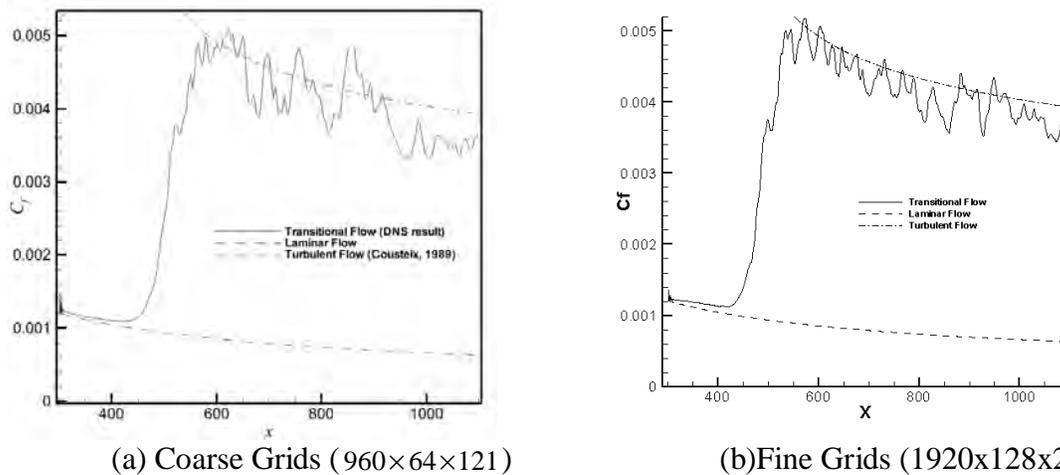

(a) Coarse Grids ($960 \times 64 \times 121$)　　　　(b)Fine Grids ($1920 \times 128 \times 241$)

Figure 5: Streamwise evolutions of the time-and spanwise-averaged skin-friction coefficient

### 2.2.3 Comparison with Log Law

Time-averaged and spanwise-averaged streamwise velocity profiles for various streamwise locations in two different grid levels are shown in Figure 6. The inflow velocity profiles at x=300.79 $\delta_{in}$ is a typical laminar flow velocity profile. At x=632.33$\delta_{in}$,the mean velocity profile approaches a turbulent flow velocity profile (Log law). This comparison shows that the velocity profile from the DNS results is turbulent flow velocity profile and the grid convergence has been realized.

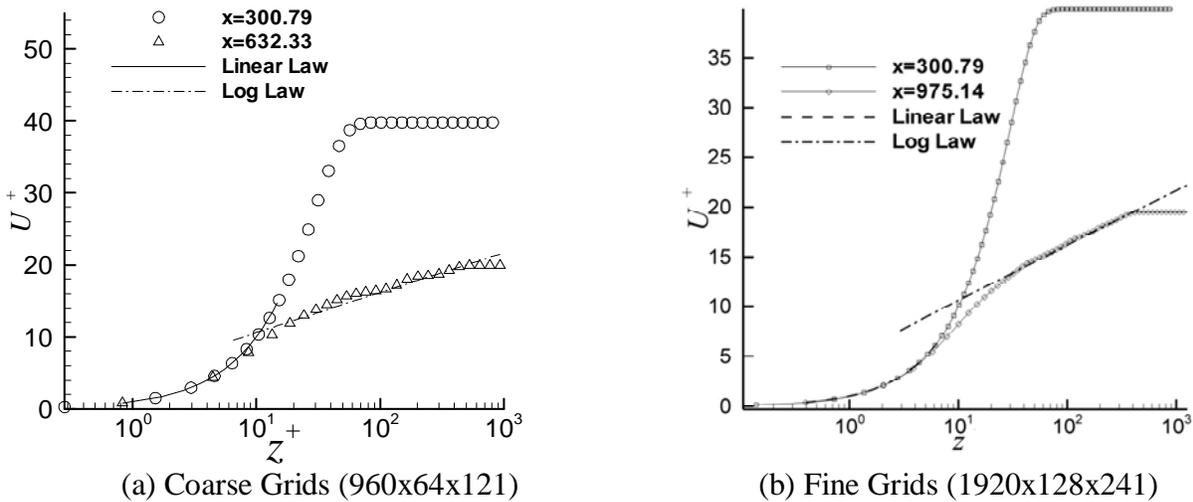

(a) Coarse Grids (960x64x121)       (b) Fine Grids (1920x128x241)

Figure 6: Log-linear plots of the time-and spanwise-averaged velocity profile in wall unit

### 2.2.4 Comparison with Other DNS

Although we cannot compare our DNS results with those given by Borodulin et al (2002) quantitatively, we still can find that the shear layer structures are very similar in two DNS computations in Figure 7.

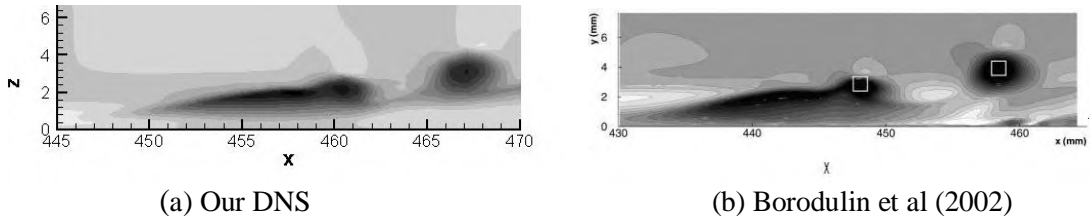

(a) Our DNS       (b) Borodulin et al (2002)

Figure 7: Qualitatively Comparison of contours of streamwise velocity disturbance u in the (x, z)-plane (Light shades of gray correspond to high values)

### 2.2.7 U-shaped vortex in comparison with experimental results

Figure 8(a) (Guo et al, 2010) represents an experimental investigation of the vortex structure including ring-like vortex and barrel-shaped head (U-shaped vortex). The vortex structures of the nonlinear evolution of T-S waves in the transition process are given by DNS in Figure 8(b). By careful comparison between the experimental work and DNS, we note that the experiment and DNS agree with each other in a detailed flow structure comparison. This cannot be obtained by accident, but provides the following clues: 1) Both DNS and experiment are correct 2) Disregarding the differences in inflow boundary conditions (random noises VS enforced T-S waves) and spanwise boundary conditions (non-periodic VS periodic) between experiment and DNS, the vortex structures are same 3) No matter K-, H- or other types of transition, the final vortex structures are same 4) There is an universal structure for late boundary layer transition 5) turbulence has certain coherent structures (CS) for generation and sustenance.

**Coherent structure of U-shaped vortex**

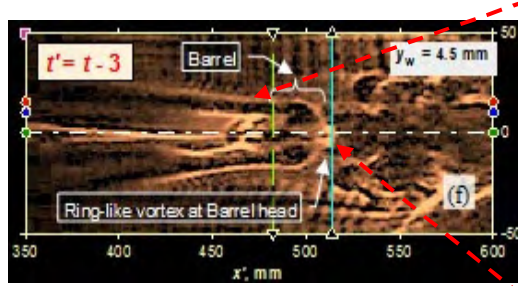

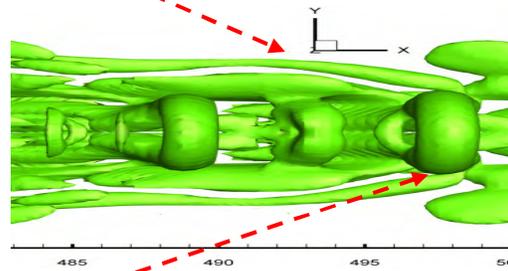

(a) Experiment results given by Guo et al (2010) vortex

(b) DNS result of U-shaped

**Ring-like vortex**

Figure 8: Qualitative vortex structure comparison with experiment

**All these verifications and validations above show that our code is correct and our DNS results are reliable.**

## III. Our DNS Observations on "Randomization"

### 3.1 Nature of coherent structure in late stage of transition

To gain additional insight for flow randomization process in late stage of transition, here we present short review of development and evolution of coherent vertical structures shaped by interaction of non linear T-S wave in the late stage. For identification purpose, we use $\lambda_2$-eigenvalue technology developed by Jeong and Hussain (1995).

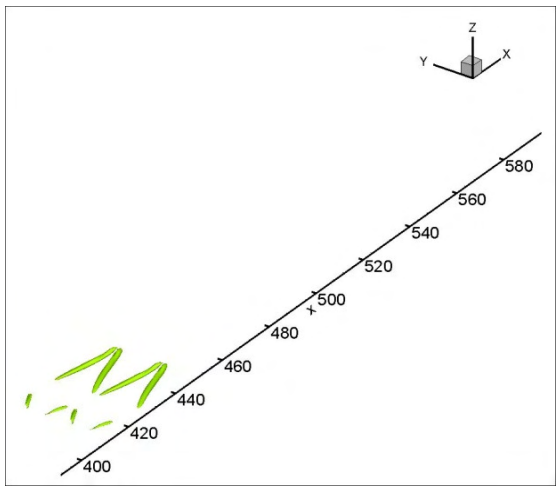

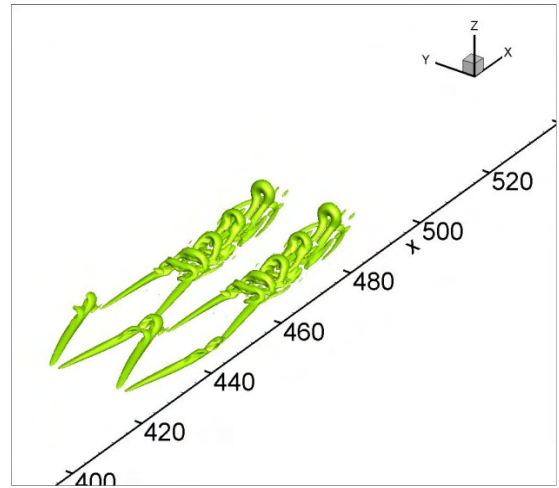

(a) t = 6T

(b) t = 7T

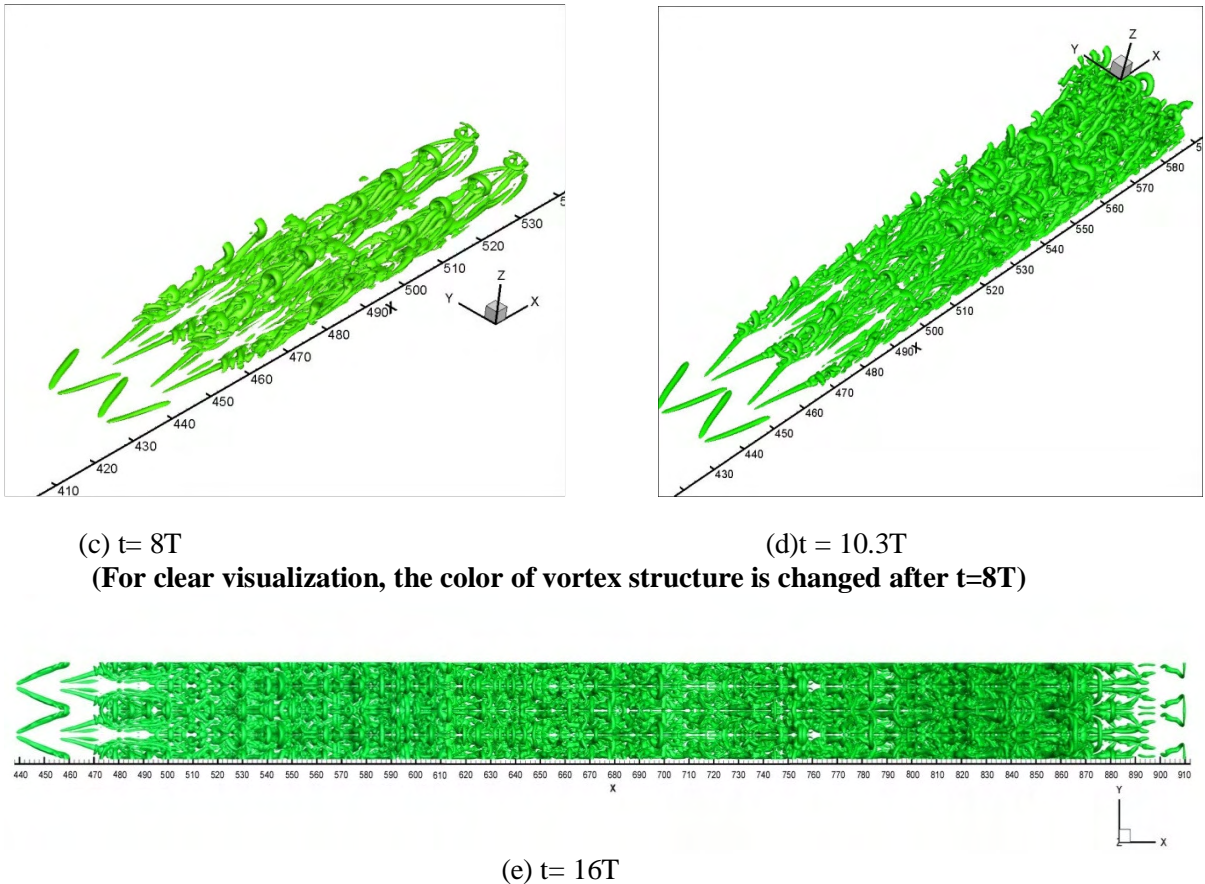

(c) t= 8T  (d)t = 10.3T

**(For clear visualization, the color of vortex structure is changed after t=8T)**

(e) t= 16T

Figure 9: Evolution of vortex structure at the late-stage of transition
(Where T is the period of T-S wave)

Late stage of transition starts with development of Λ(horse shoe)-vortices in time t=6T. These structures are rather short at the beginning (x=412-420) Figure 9(a). They are stretching continuously during evolution and become much larger while moving downstream. While moving downstream furthermore, the vortex tips are reconnected and Ω-(hairpin) vortices appear. Perfectly circular and perpendicular ring like vortices are generated by the interaction of primary and secondary streamwise vortices and they are gradually lifted up due to boundary layer mean velocity profile. Two important phenomena in late stage, namely "sweep" and "ejection" are connected with ring like vortices. The sweep between legs of ring like vortex brings low speed flow from boundary layer bottom to high speed zone in inviscid area. So, high shear layer is formed just above the ring legs. This shear layer is very unstable and multiple rings are formed by following first Helmholtz vortex conservation law (Figure 9b). For detail mechanism [6,7,18]. From Figure 9(c), we observe that second ring cycle overlaps first cycle (x= 472 to 490). This phenomenon will be described in more detail in section (3.3). The coherent vertical structures which were demonstrating very salient feature at the beginning of late stage, now started to entangle each other. As we see from Figure 9(d), the third level cycle just starts to overlap previous cycles in time step t=10.3T at x≈500. The complicated and nonlinear flow field in late stage of time can be visualizing from Figure 9(e).

## 3.2 Asymmetric phenomenon starts from second ring cycle in middle of both streamwise and spanwise direction

Here we consider the flow field which is viable through our DNS. Fig 10 (a) and (b) are top and bottom view of vortices structure at t=16.25T. We observe that both top ring and bottom ring structures are symmetric. Mean while, a slice from Figure 10(a) is chosen in stremwise direction at x=838.9 to investigate the mechanism of randomization in spanwise direction. Streamtraces are helpful to check intensity of vortices.

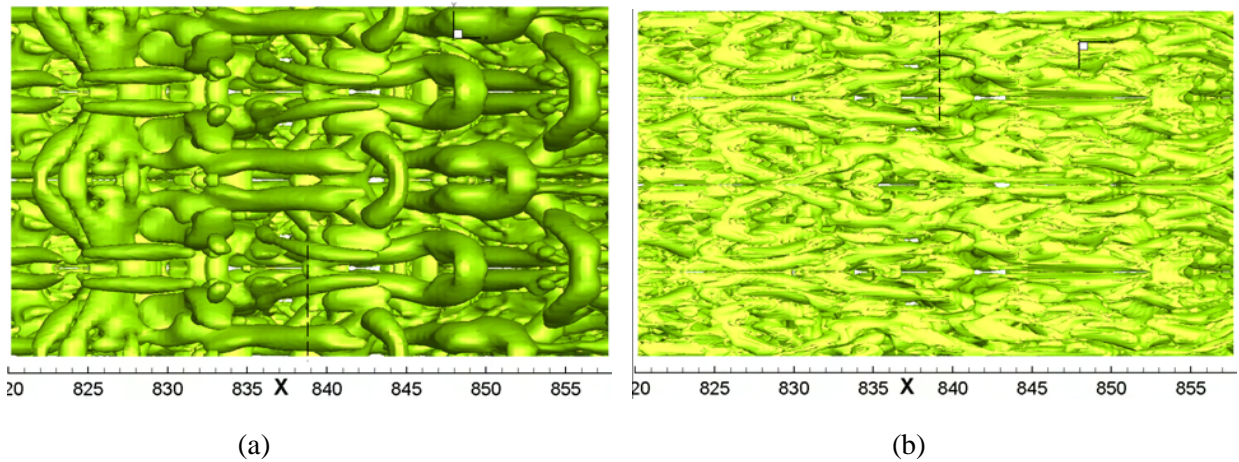

(a)                                      (b)

Figure 10:  top and bottom view of $\lambda_2$

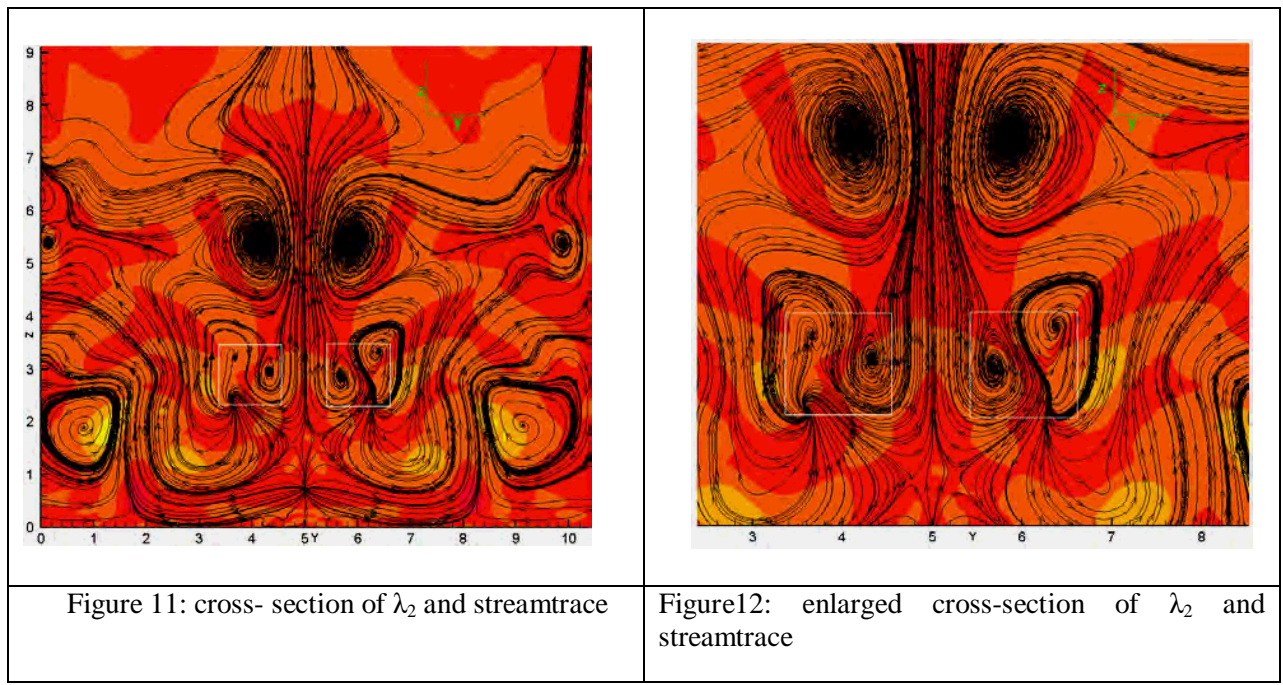

Figure 11: cross- section of $\lambda_2$ and streamtrace

Figure12:  enlarged cross-section of $\lambda_2$ and streamtrace

We can clearly see from Figure 11 that the two vortex rings inside left and right white rectangular boxes are generated with different intensity. However, all other vortices have same magnitude of intensity. To justify that the two vortices indeed have different intensity, we further check two other variables namely pressure and spanwise vorticity ($\omega_x$) at the same position (where these left and right vortices are located) in the same slice. From Figure 13, we see that there is high pressure at the left vortex ring position than in the right. Also, from Figure 14, we can visualize that left vortex ring has greater intensity of spanwise vorticity than the left vortex ring.

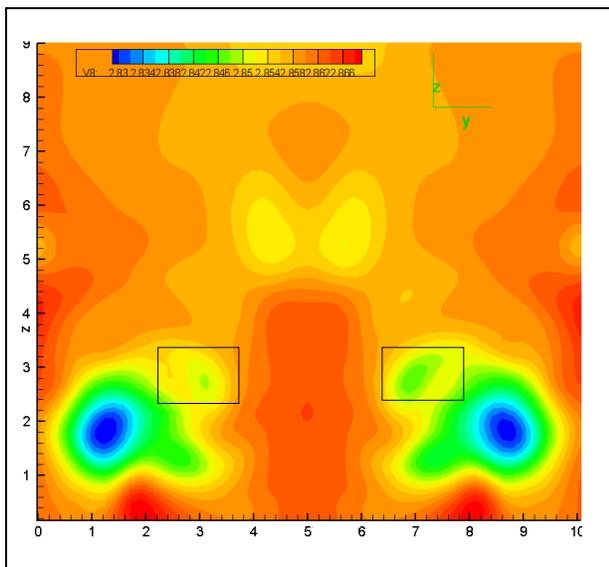

Figure 13 : cross-section of pressure

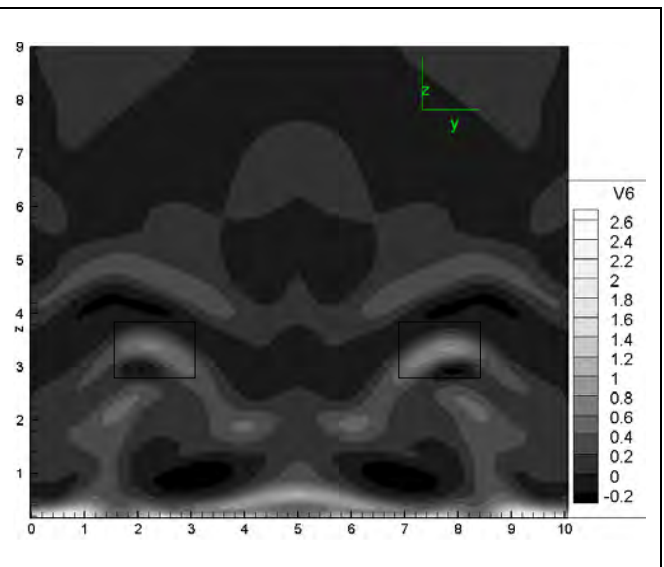

Figure 14: cross-section of spanwise vorticity

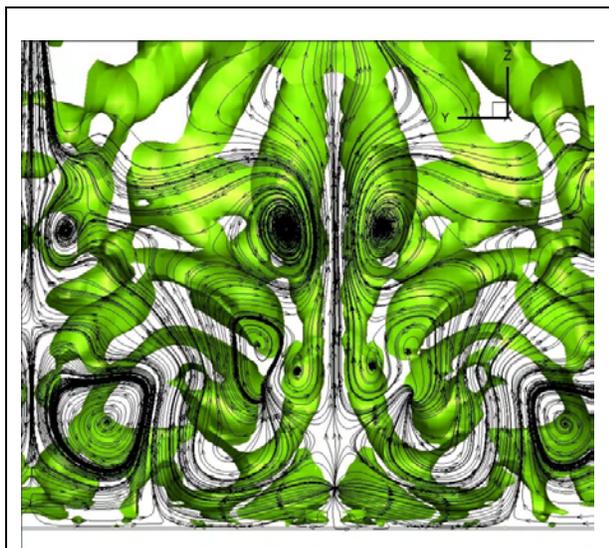

Figure 15: tail view of isosurface of $\lambda_2$

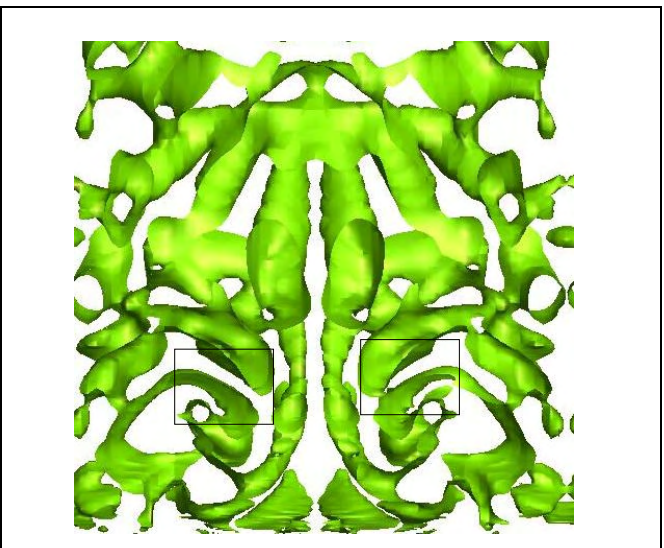

Figure 16 : enlarged  tail view of isosurface of $\lambda_2$

As we know that $\lambda_2$ technology is one of the convincing methods for visualization, we further try to confirm our claim (3.2) by using 3-dimensional vertical structures.  Figure 15 is tail view of isosurface of $\lambda_2$ at t=16.25T. Here, we cut our domain in such a way that the visible tail of vertical structure starts from

American Institute of Aeronautics and Astronautics

exactly the same slice in Figure 11 and 12. From the black rectangles in Figure 16, it can be clearly observe that the vortex structures are not symmetric.

## 3.3 Overlapping of multiple ring cycles

Figure 17 is side view of isosurcface of $\lambda_2$ at t=16.25T. From this Figure, we observe that the transitional boundary layer is getting thicker and thicker. This thickening is due to overlapping of multiple ring cycles. This overlapping phenomenon can be described in this way. Since, the ring head is located in the inviscid area and has much higher moving speed than the ring legs which are located near the bottom of the boundary layer, the hairpin vortex is stretched and multiple rings are generated. This will lead to an overlapping of second ring cycle upside of first ring cycle. However, no mixing of two cycles is observed by our new DNS (Figures 9c and 9d). The second ring is isolated from the first ring cycle by a secondary group of rings which are generated by the wall surface, separated from wall and convected to downstream. By the same reason the third ring cycle overlaps first two cycles and so on. This is the reason why the transitional boundary layer becomes thicker and thicker.

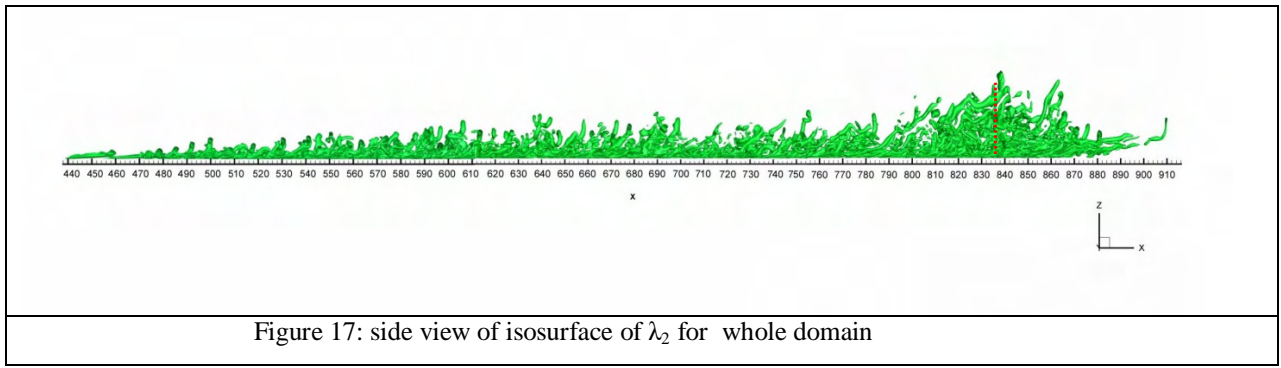

Figure 17: side view of isosurface of $\lambda_2$ for whole domain

While investigating the mechanism for randomized flow, we found an interesting connection between the origin of randomization (x=838.9) and thickness of transitional boundary layer. From fig (10), we can see that the randomization starts from the place where the boundary layer has maximum thickness. We found that the loss of symmetry happens in the middle of the flow field in the streamwise direction, not inflow and not outflow. Since all noises are mainly introduced through the inflow, outflow or far field, it is unlikely that the reason to cause asymmetry is due to the large background noises, but is pretty much the internal property of the multiple level vortex structure in boundary layers.

## 3.4 Completely asymmetric flow at very late stage

As we found the asymmetric fluctuation is originated at middle level ring cycle. Another immediate question may be how this loss of symmetry spreads in normal direction. To answer this question, we investigate the coherent vertical structures at t=17.625T.

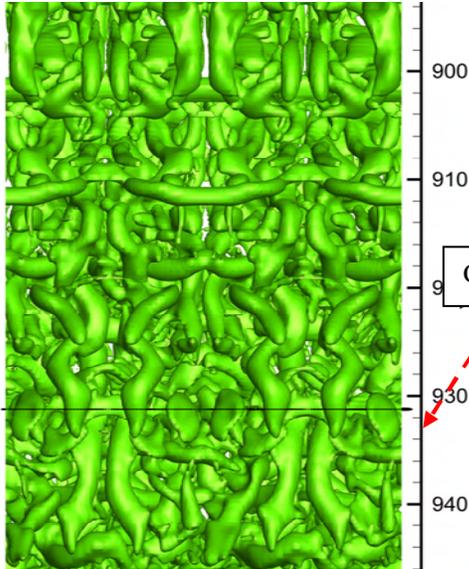

Figure 18(a). Top view isosurface of $\lambda_2$

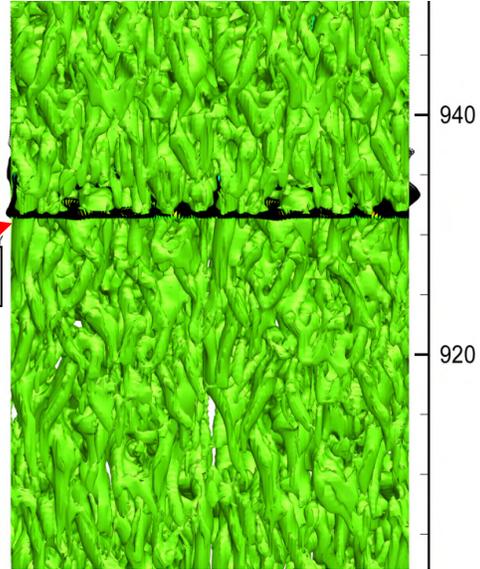

Figure 18(b). Bottom view - isosurface of $\lambda_2$

By observing Figure 18(a), we notice the top level rings still preserve asymmetry. Mean while we found that the bottom level of ring cycles completely lost symmetry (Figure 18b). This can be concluded that the asymmetric phenomenon which was started in middle just affected bottom level ring cycle. Here we found "Sweep" motion play important role to spread the asymmetric phenomenon to bottom. Actually, the sweep brings high speed fluid from inviscid region to bottom of boundary layer. The small length scale in bottom is direct consequence of sweep so these small scales are now victim. To further confirm the claim that the deformed vortices in middle affect the small scale on bottom, we choose a cross section in fig 18 (a) and (b) at x≈931.5 and draw stracetraces along each vortex (Figure 19).

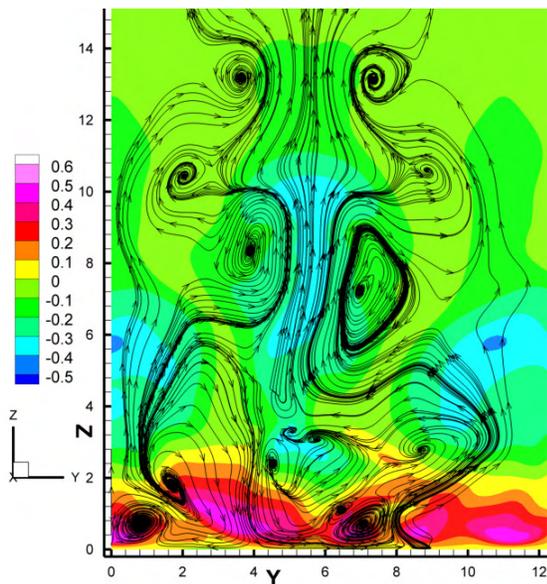

Figure 19: velocity perturbation with streamtraces

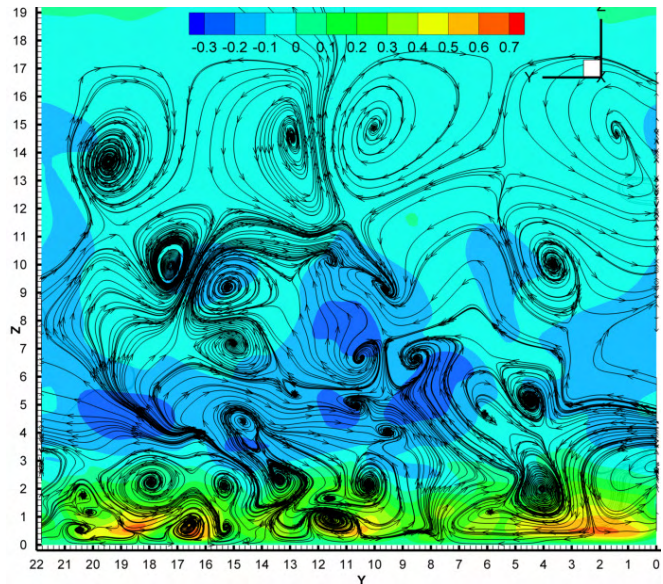

Figure 20: velocity of streamtracs

We observe most of vortices in middle and bottom are generated by different level of shear layer. Fig is top view of isosurface of where we see most top level rings are also asymmetrc. Meanwhile we select a cross-section from Figure 21, we found  almost all the flow field is asymmetric.

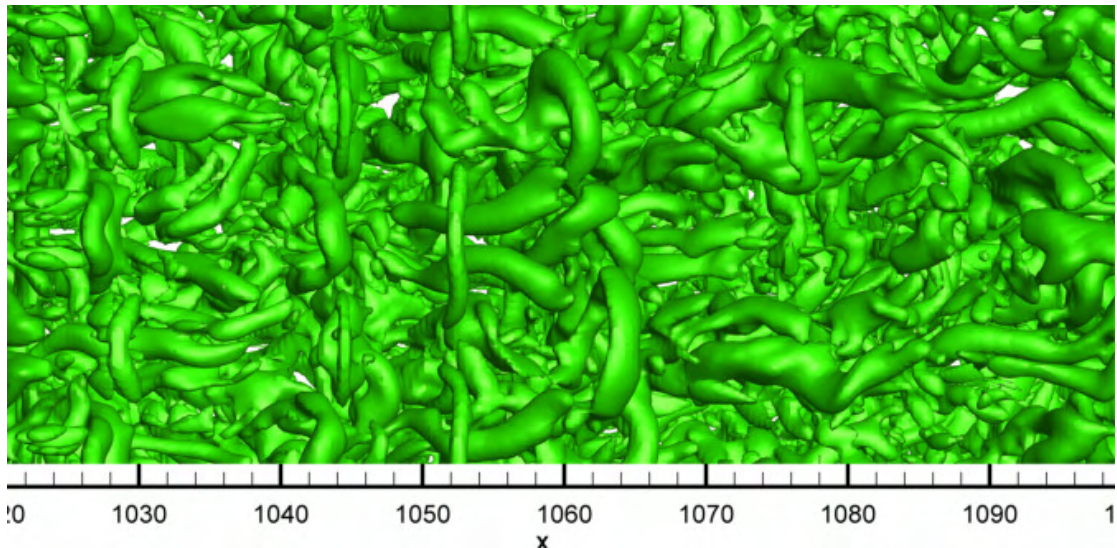

Figure 21: Top view of isosurface of $\lambda_2$

**IV. Analysis**

The flow is symmetric and periodic with period=$\pi$ (Figure 22) at early stages. Since we enforce periodic boundary condition in the spanwise direction with a period=$2\pi$ and enforce a symmetric and periodic (period=$\pi$) flow profile at the inflow, the flow will have to keep periodic in the spanwise direction (period=$2\pi$) and symmetric at inflow. However, the following processes have been observed:

1) Flow lost symmetry in the middle (not at inflow or outflow) in the streamwise direction and the middle of the multiple overlapping ring cycles (Figure 23). However, the flow is still periodic with a period=$\pi$ (Figures 24(a) and 24(b)). This means the flow does not only have $\sum_{k=0}^{n} a_k \cos(2ky)$ but also have $\sum_{k=0}^{n} b_k \sin(2ky)$ which is newly generated;

2) Flow lost periodicity with period=$\pi$, but has to be periodic with period=$2\pi$ (Figure 24(c)), which we enforced. Since the DNS study is focused on the mechanism of r andomization and the DNS computation only allows use two periods in the spanwise direction, we consider the flow is randomized when the symmetry is lost and period is changed from $\pi$ to $2\pi$ (Figure 25):

$$f(y) = \sum_{k=0}^{n} a_k \cos(2ky) + \sum_{k=1}^{n-1} b_k \sin(2ky) + \sum_{k=0}^{n-1} c_k \cos(2ky + y) + \sum_{k=1}^{n-1} d_k \sin(2ky + y)$$

In real flow, there is no such a restriction of periodic boundary condition in the spanwise direction.

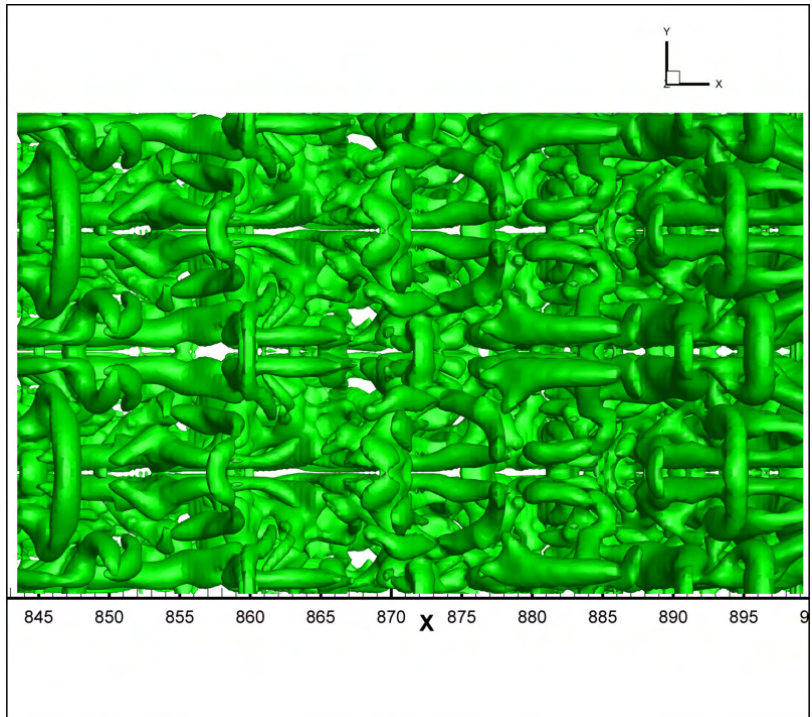

Figure 22: Whole domain is symmetric and periodic: f(-y)=f(y) and f(y+π)=f(y)
(the period=π; spanwise domain is −π<y<π)

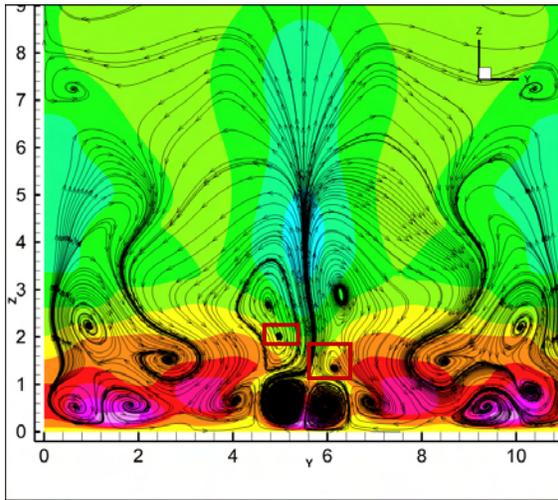

(a) Section view in y-z plane

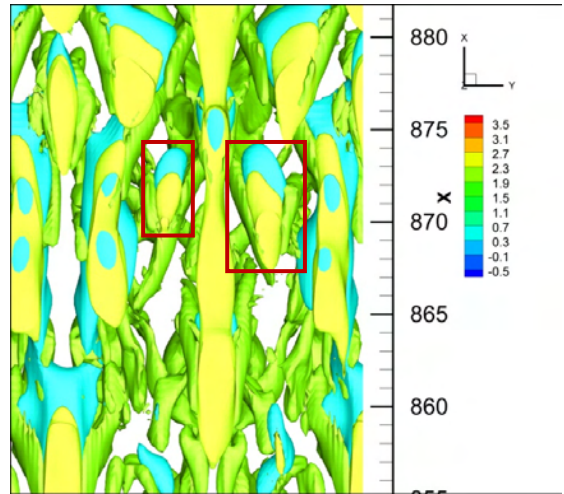

(b) Bottom view of positive spike

Figure 23: The flow lost symmetry in second level rings and bottom structure at T=15.0

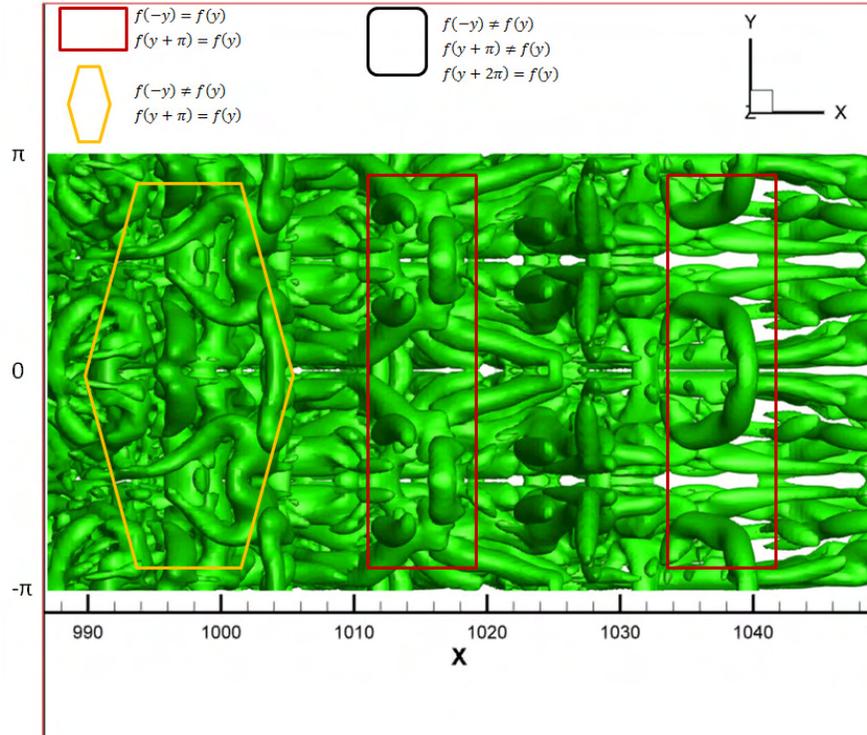

Figure 24(a): Symmetric and asymmetric – Red rectangular frame: periodic and symmetric at y=-π/2, 0, π/2, i.e. f(y+π)=f(y), f(-π/2 −y)=f(-π/2+y), f(-y)=f(y), f(π/2-y)=f(π/2+y); Yellow diamond frame: periodic, f(y+π)=f(y), period=π; but asymmetric **f(−y) ≠ f(y)**; the spanwise domain is −π<y<π

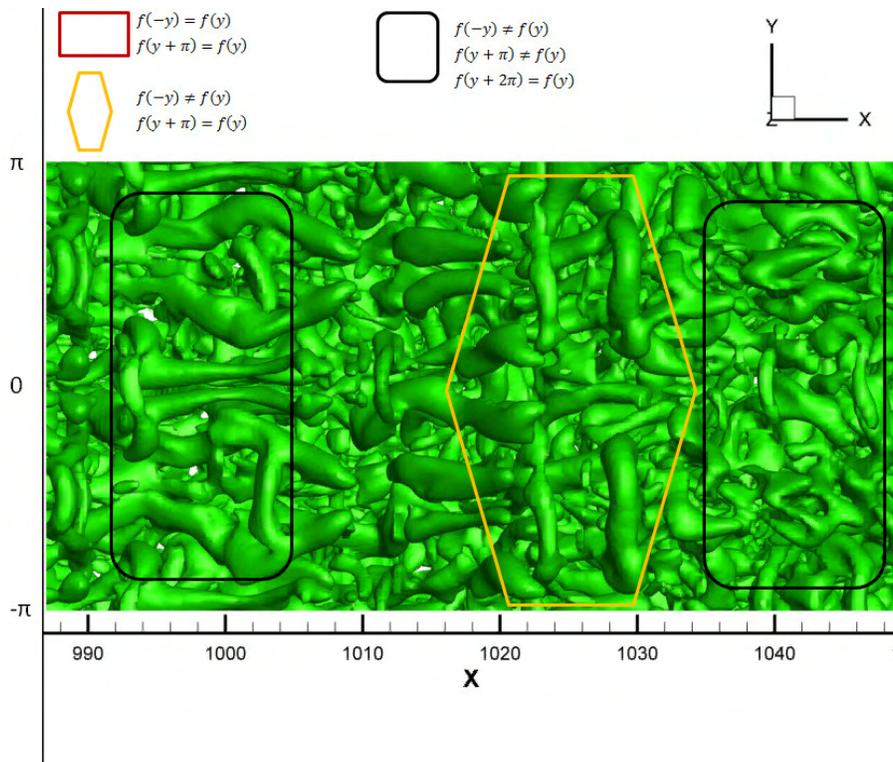

Figure 24(b): Periodic but asymmetric – Yellow diamond frame: periodic, period=π; black box: periodic but period= 2π; the spanwise domain is −π<y<π

American Institute of Aeronautics and Astronautics

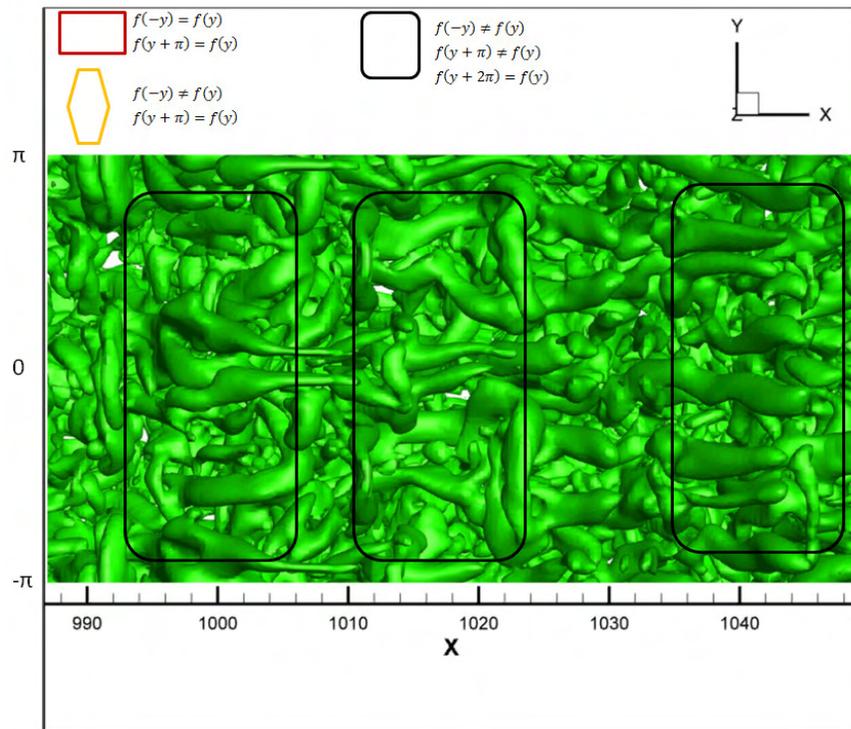

Figure 24(c): Periodic but asymmetric – all black boxes: periodic but asymmetric (period= 2π); the spanwise domain is −π<y<π

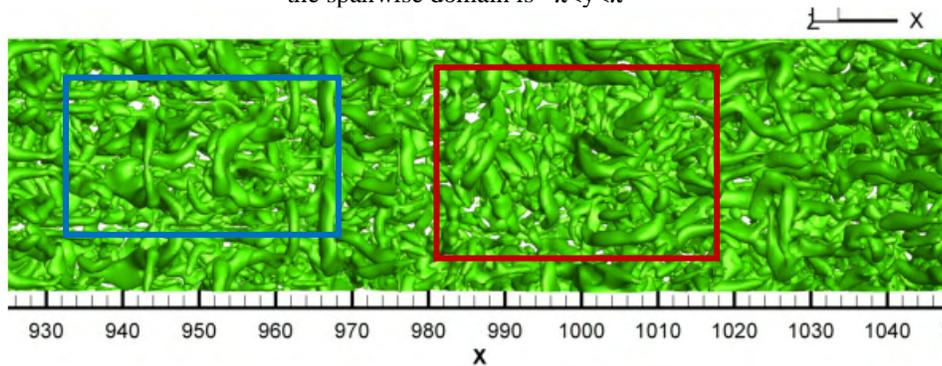

Figure 25:  The whole flow field lost symmetry at T=21.25
Top ring structure lost symmetry (blue area is symmetric but red area is not)

## V. Hypothesis

Apparently, loss of symmetry, which is equivalent to adding sin (2ky), is a key issue for the starting point of flow randomization. A hypothesis was given by Liu that the loss of symmetry is caused by the shift from C-type to K-type transitions or reverse. The randomization is caused by the instability of the multiple level vortex ring cycle structure which is an internal property. Therefore, randomization is not mainly caused by big background noises or removal of periodic boundary conditions in DNS. The flow shows a C-typed transition (staggered) at beginning (Figure 26(a)), but becomes K-type transition later (Figure 26(b)) and then mixed (Figure 26(c)). This means the first vortex circle is C-type, but second circle, which overlaps the first circle, is K-type.  K-type, C-type, and mixed type are all observed by experiment. There must be some trend of shift from C-type to K-type or reverse. This shift will cause the loss of symmetry of the middle of the first circle (Figure 27). This trend will change the underneath large ring structure and cause the loss of symmetry of the underneath large rings. Once the middle large rings lost symmetry, the underneath small length scale will lose the symmetry quickly due to the asymmetry of second sweeps. Finally, the asymmetry of lower level vortex structures will affect the top rings through ejection and the whole flow field becomes turbulent.

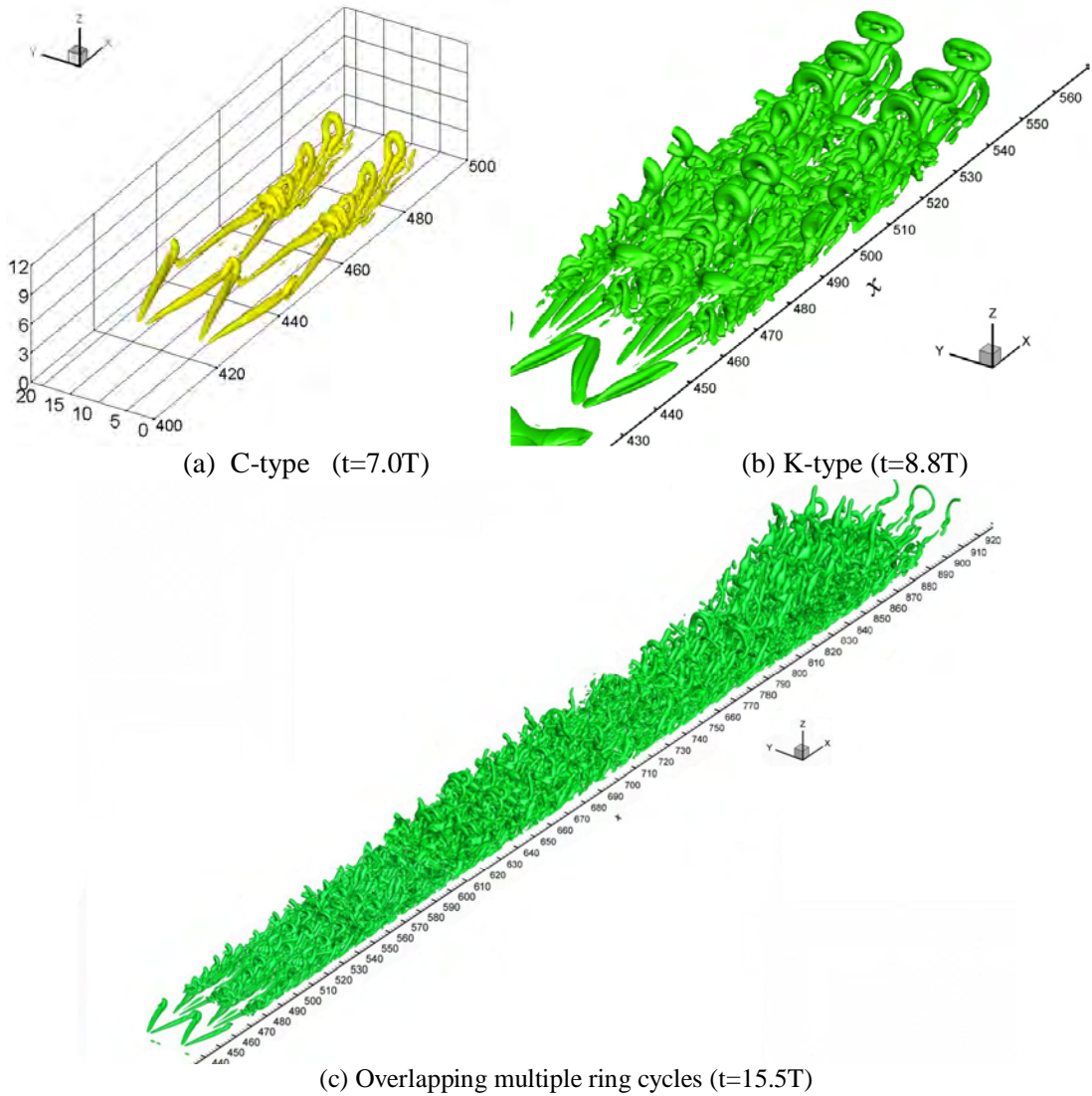

(a) C-type (t=7.0T)    (b) K-type (t=8.8T)

(c) Overlapping multiple ring cycles (t=15.5T)

Figure 26: Vortex structure in K-type, C-type and mixed type transition (T is the T-S period)

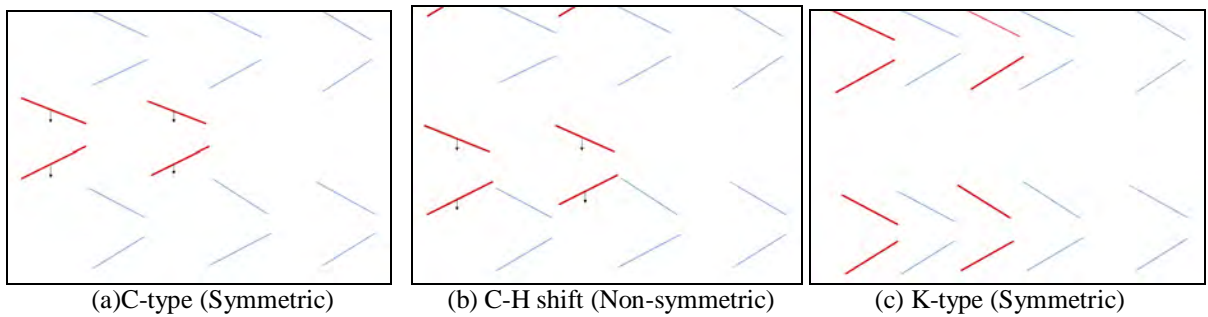

(a)C-type (Symmetric)    (b) C-H shift (Non-symmetric)    (c) K-type (Symmetric)

Figure 27: Sketch of symmetry loss due to the shift from C-type transition (vortex circles are staggered) to the K-type transition (vortex circles are aligned) (the blue one is the first circle and the red one is the second circle)

# VI. Conclusion

Although we still use the symmetric boundary condition (period is π for inflow and 2π for the whole domain) without intentional introduction of background noises from the inflow, outflow, and far-field, we still find that the vortices become asymmetric in the whole flow field finally. Following conclusions can be made by the current DNS result we obtained:

1. The phenomenon of asymmetry is first observed at the middle level of the overlapping multiple ring cycles instead of the ring tip. Meanwhile, the loss of flow symmetry is also found at the middle part of the flow field in the streamwise direction. Meanwhile, the neighboring inflow and neighboring outflow parts still keep symmetric characteristics.

2. The background noise could prompt the C-type or K-type transition. Eventually, the loss of symmetry may be caused by the shift from C-type to K-type transitions or reverse. In addition, randomization could be caused by the instability of the multiple level vortex ring cycle structure as well. Both are the internal property.

3. There are small vertex rings generated at the middle by different streamwise velocity shear levels which will affect the intensity of positive spikes. This will result in deformation of the small vortices near the bottom of the boundary layer.

4. The asymmetric lower level vortices deform the shape of the upper level vortices through ejection. Simultaneously, the deformed small vortices in middle quickly affect the small length scales in the bottom.

5. Finally, we can find that the top flow structure loses the symmetry and the whole flow field is randomized and becomes turbulent.

In summary, the order of asymmetric phenomenon can be concluded as middle ring structure first, bottom later and top last.

## Acknowledgments

This work was supported by The Department of Mathematics at University of Texas at Arlington. The authors are grateful to Texas Advanced Computing Center (TACC) for providing computation hours. This work is accomplished by using Code LESUTA which was released by Dr. Chaoqun Liu at University of Texas at Arlington in 2009.

## References:

[1]Lu,P., Thapa, M., and Liu, C. Numerical Study on Randomization in Late Boundary Layer Transition, AIAA 2012, Nashville, TN, January 2012

[2]MEYER, D.G.W.; RIST, U.; KLOKER, M.J. (2003):Investigation of the flow randomization process in a transitional boundary layer. In: Krause, E.; Jäger, W. (eds.): High Performance Computing in Science and Engineering '03. Trans¬actions of the HLRS 2003, pp. 239-253 (partially coloured), Springer

[3]Bake S, Meyer D, Rist U. Turbulence mechanism in Klebanoff transition:a quantitative comparison of experiment and direct numerial simulation. J.Fluid Mech, 2002 , 459:217-243

[4]Boroduln V I, Gaponenko V R, Kachanov Y S, et al. Late-stage transition boundary-Layer structure: direct numerical simulation and exerperiment. Theoret.Comput.Fluid Dynamics, 2002,15:317-337.

[5]Adrian, R. J., Hairpin vortex organization in wall turbulence, Physics of Fluids, Vol 19, 041301, 2007

[6]Chen L., Tang, D., Lu, P., Liu, C., Evolution of the vortex structures and turbulent spots at the late-stage of transitional boundary layers, Science China, Physics, Mechanics and Astronomy, Vol. 53  No.1: 1□14, January 2010b,

[7]Chen, L., Liu, X., Oliveira, M., Tang, D., Liu, C., Vortical Structure, Sweep and Ejection Events in Transitional Boundary Layer, Science China, Series G, Physics, Mechanics, Astronomy,  Vol. 39 (10)

[8]Chen, L., Liu, X., Oliveira, M., Liu, C., DNS for ring-like vortices formation and roles in positive spikes formation, AIAA Paper 2010-1471, Orlando, FL, January 2010a.

[9]Guo, Ha; Borodulin, V.I..; Kachanov, Y.s.; Pan, C; Wang, J.J.; Lian, X.Q.; Wang, S.F., Nature of sweep and ejection events in transitional and turbulent boundary layers, J of Turbulence, August, 2010

[10]Jeong J., Hussain F. On the identification of a vortex, J. Fluid Mech. 1995, 285:69-94

[11]Duros F, Comte P, Lesieur M. Large-eddy simulation of transition to turbulence in a boundary layer developing spatially over a flat plate. J.Fluid Mech, 1996, 326:1-36

[12]Jiang, L., Chang, C. L. (NASA), Choudhari, M. (NASA), Liu, C., Cross-Validation of DNS and PSE Results for Instability-Wave Propagation, AIAA Paper #2003-3555, The 16th AIAA Computational Fluid Dynamics Conference, Orlando, Florida, June 23-26, 2003

[13]Liu, X., Chen, L., Oliveira, M., Tang, D., Liu, C., DNS for late stage structure of flow transition on a flat-plate boundary layer, AIAA Paper 2010-1470, Orlando, FL, January 2010a.

[14]Kachnaov, Y. S., 1994, "Physical Mechanisms of Laminar-Boundary-Layer Transition," Annu. Rev.Fluid Mech., 26, pp. 411–482.

[15]Lu, P. and Liu, C., DNS Study on Mechanism of Small Length Scale Generation in Late Boundary Layer Transition, AIAA Paper 2011-0287 and J. of Physica D, Non-linear, to appear, 2011b, on line: http://www.sciencedirect.com/science/article/pii/S0167278911002612

[16]Rist, U. and Kachanov, Y.S., 1995, Numerical and experimental investigation of the K-regime of boundary-layer transition. In: R. Kobayashi (Ed.) Laminar-Turbulent Transition (Berlin: Springer) pp. 405-412.

[17]Wu, X. and Moin, P., Direct numerical simulation of turbulence in a nominally zero-pressure gradient flat-plate boundary layer, JFM, Vol 630, pp5-41, 2009

[18]Chen, L and Liu, C., Numerical Study on Mechanisms of Second Sweep and Positive Spikes in Transitional Flow on a Flat Plate, Journal of Computers and Fluids, Vol 40, p28-41, 2010

[19]Schlichting, H. and Gersten, K., Boundary Layer Theory, Springer, 8th revised edition, 2000

[20]Kolmogorov, Andrey Nikolaevich (1941). "The local structure of turbulence in incompressible viscous fluid for very large Reynolds numbers". Proceedings of the USSR Academy of Sciences 30: 299–303. (Russian), translated into English by Kolmogorov, Andrey Nikolaevich (July 8, 1991). "The local structure of turbulence in incompressible viscous fluid for very large Reynolds numbers". Proceedings of the Royal Society of London, Series A: Mathematical and Physical Sciences 434 (1991): 9–13. Bibcode 1991RSPSA.434....9K. doi:10.1098/rspa.1991.0075.

[21]Liu.,C. Numerical and Theoretical Study on "Vortex Breakdown", International Journal of Computer Mathematics, (to appear)

[22]Kachnaov, Y. S., 1994, "Physical Mechanisms of Laminar-Boundary-Layer Transition," Annu. Rev. Fluid Mech., 26, pp. 411–482.